\documentclass[sigconf]{acmart}

\AtBeginDocument{%
  \providecommand\BibTeX{{%
    \normalfont B\kern-0.5em{\scshape i\kern-0.25em b}\kern-0.8em\TeX}}}

\setcopyright{acmcopyright}
\copyrightyear{2023}
\acmYear{2023}
\acmDOI{XXXXXXX.XXXXXXX}

\acmConference[CBMI '23]{International Conference on Content-based Multimedia Indexing}{September 20--22,
  2023}{Orleans, France}
\acmPrice{15.00}
\acmISBN{978-1-4503-XXXX-X/18/06}

\begin{document}

\title{Handwriting Analysis on the Diaries of Rosamond Jacob}

\author{Sharmistha S. Sawant}
\author{Saloni D. Thakare}
\affiliation{%
    \institution{School of Computing, Dublin City University} 
    \streetaddress{Glasnevin}
    \city{Dublin 9}
    \country{Ireland}}

\author{Derek Greene}
\affiliation{
    \institution{School of Computer Science, \\University College Dublin}
    \streetaddress{Belfield}
    \city{Dublin 4}
    \country{Ireland}
}

\author{Gerardine Meaney}
\affiliation{
    \institution{School of English, Drama and Film,\\University College Dublin}
    \streetaddress{Belfield}
    \city{Dublin 4}
    \country{Ireland}
    }

\author{Alan F. Smeaton}
\affiliation{%
    \institution{Insight SFI Research Centre for Data Analytics,\\Dublin City University} 
    \city{Dublin 9}
    \country{Ireland}}
    \email{alan.smeaton@dcu.ie}

\renewcommand{\shortauthors}{Sawant, et al.}

\begin{abstract}
Handwriting is an art form that most people learn at an early age. Each person's writing style is unique with small changes as we grow older and  as our mood changes. Here we  analyse  handwritten text in a culturally significant personal diary. We compare  changes in  handwriting and  relate this to the sentiment of the  written material and to the topic of  diary entries. We identify  handwritten text from  digitised images and  generate a canonical form for  words using shape matching  to compare how the same handwritten word appears over a period of time. For determining the sentiment of  diary entries, we use the Hedonometer, a dictionary-based approach to scoring sentiment. 
We apply these techniques to the historical
diary entries of Rosamond Jacob (1888-1960), an Irish writer and political activist whose daily diary entries report on the major events in Ireland during the first half of the last century.
\end{abstract}




\maketitle

\section{Introduction}

Journaling  is  a  process of  writing  one's  daily life  experiences  and    recording    unforgettable and forgettable  moments and everyday activities \cite{10.1145/2858036.2858103}. Journaling in different forms has been practised for centuries and journals can provide insights into an individual as well as into the lifestyle, habits and happenings during the era that a journal was written.   
Rosamond Jacob was an Irish activist and suffragist who wrote a daily journal which paints  a picture of  political and daily Irish life up to 100 years ago. Her diary entries are legible, detail-oriented, and preserved by the National Library of Ireland\footnote{\url{https://catalogue.nli.ie/Collection/vtls000029815/CollectionList}}. They record the contributions of women who changed the political scene in Ireland a century ago. A collection of the diary entries are digitised as high-quality scanned images, with corresponding transcripts created by a literary expert. Rosamond Jacob’s diaries are a unique and extraordinarily rich resource for the understanding of twentieth century Irish history, especially in relation to the role of women, which have been under-utilised as a source of firsthand testimony due to the difficulty of searching volumes of handwritten notebooks for specific topics.

Every individual has a unique style of handwriting and there are only small changes to our handwriting style as we grow older \cite{10.1007/978-3-031-09037-0_24}. 
Previous work has shown how handwriting is also linked to the mood of the writer at the time of writing, which is in turn influenced by the topic of that handwriting \cite{likforman15handwriting,rispler2018mood}.  This means that the handwriting in journal entries may reveal insights into the writer's mood when describing certain topics, events or people. 

In this paper we present a study on the diaries of Rosamond Jacob to determine if we can detect changes in her handwriting style over time and whether such changes might correlate either with the writer's emotional state or the topics of her personal diary entries at a given point in time. 

\section{Background}

Every individual has a different handwriting style which evolves slowly over time. Handwriting includes characteristics such as stroke, slant, line spacing, and the shapes of the characters. These are the key features used in statistical methods for handwriting pattern matching  \cite{adamek2007word}, either when considering a single writer or when making comparisons between multiple different writers  \cite{srihari2001individuality}. Artificial neural networks have been used for classification of handwriting, where results show 98\% accuracy for the task of identifying a writer, indicating that these features are robust and useful \cite{8620507}.

When detecting handwritten text from scanned images, it is important to  identify the page orientation and  lines on which the text is written, either real  or imaginary ruler lines. The  detection and extraction of text from curved scanned images, especially when the pages are bound into book format, was previously considered in \cite{shejwal2017segmentation}. Typically this task proceeds as follows. Firstly, images are converted to grayscale format and then binarised. Removal of noise such as dirt or stains on the paper can be done by morphological image processing. Subsequently, 
 segmentation of words requires smoothing of text lines using a Gaussian filter. A bounding box is created around each word and text line segmentation is achieved using an approach such as the $k$-nearest neighbour algorithm with Euclidean distance.

{\bf Canonical Correlation Analysis.} 
Multi-view data can be found in many real-world scenarios where the same data can be seen from different perspectives. Canonical Correlation Analysis (CCA) \cite{hardoon04cca} is a common approach  used in multi-view learning to merge or fuse multiple views of a single point of data.
An overview of different approaches to multi-view learning is given in \cite{guo2019canonical}. Traditionally, CCA is  unsupervised and uses only two views to calculate the maximum correlation between them. However, in real world data, the relationship between two views can be non-linear and an alternative method called Kernel CCA finds the non-linear correlation between two views of data. 

Deep Canonical Correlation Analysis (DCCA) uses deep neural networks to understand the non-linearity between two views and uses the linear method to find the canonical correlations \cite{andrew2013deep}. Previous works using both MNIST and speech data
have shown that DCCA outperforms previous CCA  approaches to learning representations from a dataset. In an attempt to generalise the canonical correlation theory, tensor CCA \cite{luo2015tensor} was developed that maximises the correlation of multiple views of data with the help of a high-order covariance tensor. This  is used in various multi-view recognition problems such as handwritten digit recognition and face recognition.  
In this paper we use correlation analysis to create the canonical form of a multiply occurring handwritten word in a document. When a diary entry has multiple occurrences of the same word, we combine all the handwritten occurrences to generate the canonical form for that document, for subsequent analysis.

{\bf Structural Similarity Index Measure.}
To calculate  similarities among handwritten words represented as images, we consider the use of the Structural Similarity Index Measure (SSIM).  This measure incorporates a range of   edge detection techniques \cite{raju2017comparative}, including Laplacian \cite{Laplacian}, Sobel \cite{kanopoulos1988design}, Robert \cite{chaple2015comparisions} and Prewitt \cite{chaple2015comparisions}.
While alternative image similarity measures exist, such as the
peak signal-to-noise ratio (PSNR) and mean squared error (MSE), these have been shown to  be inconsistent with human visual perception. MSE and PSNR estimate absolute errors, whereas SSIM is a perception-based model that considers image degradation as perceived changes in structural information by incorporating important perceptual phenomena, including  luminance masking and contrast masking \cite{raju2017comparative}. SSIM produces values in the range $[0,1]$, where 1 represents a perfect match and 0 indicates no similarity.

{\bf Word Spotting using Dynamic Time Warping.}
Analysting historical handwritten documents requires  word spotting which groups similar looking words into clusters using  measures of image similarity. Previously, the technique of Dynamic Time Warping (DTW) \cite{senin08dtw} was applied to the George Washington document collection to build a search index \cite{rath2007word}. Each word occurrence was segmented into an image and images were clustered using DTW with each cluster labelled.  A search index was then built using  cluster labels and different similarity measures were considered for calculating word (image) similarity, with DTW showing the best performance. 

{\bf Sentiment Analysis.} 
Applying sentiment analysis to texts such as  diaries, can provide us with a  a different perspective on the narrative, and perhaps reveal some insights over time \cite{jockers20sentiment}. 
A wide variety of sentiment analysis tools have been developed, which make use of different representations, including lexical and syntactic features. However, the accuracy, reliability, and domain adaptability of many such systems is questionable  \cite{10.1145/3532203}. Common text features used in this context include part of speech tags, uni-grams, bi-grams, and surface forms of words. More recently, a theory of latent semantic analysis which applies statistical methods to learn contextual usage of words was reported in \cite{jianqiang2018deep}. This was used to find patterns in texts and combines extracted word embedding with n-grams and word sentiment polarity scores to generate a sentiment feature set. GLOVE pre-trained word embeddings \cite{pennington2014glove} were subsequently used to improve  model performance. The study concluded that using a deep CNN architecture was more effective in uncovering semantics in text data, when compared to other state-of-the-art approaches.

{\bf The Rosamond Jacob Diaries.}
\label{sec:diaries}
Rosamond Jacob (1888--1960) born in Waterford, Ireland, was an Irish writer and political activist.  She belonged to the generation who lived through two World Wars and was a protester in the Cold War nuclear threat.  Rosamond Jacob's parents were nullifidian as a result of which she lived a lonely childhood which is noticeable in her writings. She was involved in politics and was an activist for women's rights and against animal cruelty, and made her living through her writings. 

Jacob  documented all her lifelong experiences from friendships to relationships and her erudite and poetic efforts into  handwritten diaries. These  attest to what it was like to live through major  events in Ireland over a period of more than half a century. 

A collection of  171 of her diaries is held by the National Library of Ireland (NLI) in digital form\footnote{NLI Jacob diaries MS 32,582/1-171. See \url{https://www.nli.ie}}. The first 170 are dated from her early childhood (1897) until her death (1960) and the last  ((NLI MS 32,582 (171)) is her personal diary from July 1915 to August 1930.
These  have been fully transcribed by Maria Mulvany at UCD as part of a UCD Decade of Centenaries project which sought to increase public understanding of womens' roles in this key period in Irish history (Women’s Life Writing, Adam Matthews Digital, forthcoming).\footnote{These have been generously shared with the authors by Maria Mulvany.}.

It has previously been shown that handwriting is linked to the mood of the writer and the subject on which they are writing \cite{rispler2018mood} and there is potential to analyse the emotional state of a writer based on changes in their handwriting  \cite{likforman15handwriting}. In this paper we investigate whether there is a correlation between the characteristics of the handwriting and the sentiment and he topics of  diary entries as Rosamond Jacob writes about named entities which are important to her.
%
Investigating this  could lead to insights into 
the impact of pivotal moments in Irish history on ordinary activists like Jacob, who participated in many key progressive organisations without every attaining leadership positions. Using handwriting and sentiment analysis in this way offers the tantalising prospect of a glimpse into the mood and atmosphere of an historical moment and dynamic changes in that mood and atmosphere. In Jacob's case, this could provide insight into the processes by which a political and cultural activist came to actively support  violent rebellion and participate in a civil war, despite a strong pacifist background. Tracking changes in handwriting and emotion in this context offers clues to the process by which action becomes imperative in specific historical conditions where the political status quo comes to be experienced as unbearable. Given Jacob's very conscious feminism,  understanding sentiment fluctuation in the diaries has significant potential for understanding the role of gender in these complex historical process.

\section{Image Processing }
\label{sec:image}

We describe the sequence of steps  performed on a subset 1916-1920 of the Jacob diaries.
Our first step was to crop large borders  for better image processing. 
For handwriting analysis we chose all diary images from the month of October for each of the years 1917--1921 and grouped these into weeks. The length of each weekly entry varied from a single diary page to 8 pages of handwritten text.  
%

\begin{figure*}[ht]
\includegraphics[width=\linewidth]{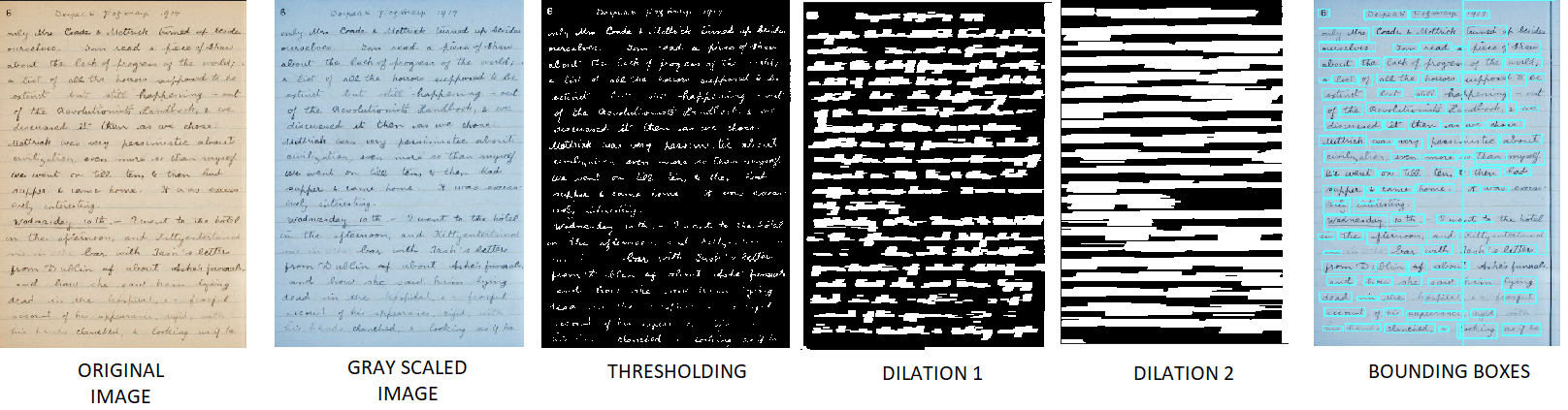}
\caption{Image processing and word segmentation steps, as applied to a sample Jacob diary page.\label{fig:preprocess}}
\end{figure*}

Using OpenCV
we performed pre-processing  namely re-sizing, grayscaling, thresholding, and dilation.
%
For re-sizing we achieved uniform proportionate size by capping the image width at 1000 pixels and adjusting the image height in proportion. 
For grayscaling
we converted   colour images using  colour space transformation methods available in  OpenCV.
Dilation is a morphological operation that processes images based on shapes. Morphology helps to extract useful components of an image either by applying dilation or erosion techniques. Dilation enhances the dimensions of the foreground region  whereas erosion shrinks the size of a foreground region. 
We dilated the binarised images by passing a kernel matrix with predefined dimensions and containing 1 as values through them using the dilation function in OpenCV. 
%
Finally,  image segmentation  breaks each image into segments based on some conditions. The output  is a collection of  masks where each segment is colour-coded based on the group it belongs to.  To segment a handwritten diary image into groups of word snippets, we applied  contour detection and word cropping. 
Figure~\ref{fig:preprocess} shows the steps  when taking an original diary entry image, resizing, converting to grayscale and creating a binarised image using thresholding. We also performed dilation twice, once to detect lines and the second  to detect words, followed by contour detection to create cropped images using the bounding boxes.  

\begin{figure*}[ht]
\includegraphics[width=0.4\linewidth]{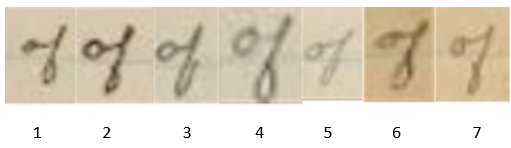}
\caption{Samples for 'of' word snippets as illustration for handwriting comparison from October 1917, week 1. This process of canonicalising the multiple occurrences of each of the 5 most common words which occur multiple time in a document, is repeated for each word.\label{of_images}}
\end{figure*}

We choose the  5 most frequently occurring words (the, of, to, in, a) from Peter Norvig's frequencies of English words from the Google Books dataset\footnote{List available at \url{http://norvig.com/mayzner.html}}.  We did not use the word ``and" because the diaries show frequent use of ``\&" instead of ``and". 
Figure~\ref{of_images} shows 7 occurrences of the word ``of"  from  diary entries in the first week of October, 1917 for illustration.  This process of canonicalising the multiple occurrences of each of these 5 words  which occur multiple time in a document, is repeated for each word.
Word snippets  cropped to different sizes  were re-scaled to the same dimensions though this resulted in some images having blurring. 

We built three models which compare word images using three image similarity metrics and built a similarity matrix among the words, using each. {\em Mean Squared Error (MSE)} which calculates the difference between two images at pixel level, squares those differences and computes the average. Mean squared error is  simple to implement but  can  mislead as larger distances in terms of pixel values does not always correspond to very different images. 
The model takes two grayscale images as input and two images are perfectly similar if the MSE value is 0 and  as the MSE value increases the images are less similar.

The {\em Structural Similarity Index (SSIM)}
compares two images with a more holistic approach rather than just at pixel level and was our second metric. SSIM does not compare pixel-to-pixel but makes comparisons between groups of pixels.  SSIM captures three aspects from an image -- Luminance, Contrast and Structure and  replicates  perceptive human behaviour when identifying similarities between  images. 
The model takes the input as two grayscaled images and two images are perfectly similar if the SSIM score is 1 and completely dissimilar if the score is 0. 

The third metric  used was {\em Dynamic Time Warping (DTW)}, an efficient algorithm to calculate the similarity between two arrays or time series with varying lengths. DTW can be applied to any data converted in a linear sequence  and uses a one-to-many matching approach to compare two unequal length arrays. 
We took grayscale images, converted them to  normalised linear sequences using the vertical projection profile method to calculate the sum of all pixel values along all vertical axes of an image. We thus represent each image as  a 1D array and calculate the DTW similarity  using the Manhattan distance to generate a similarity matrix for all  pairs of images. 


\begin{figure}[!htb]
  \centering
  \includegraphics[width=\linewidth]{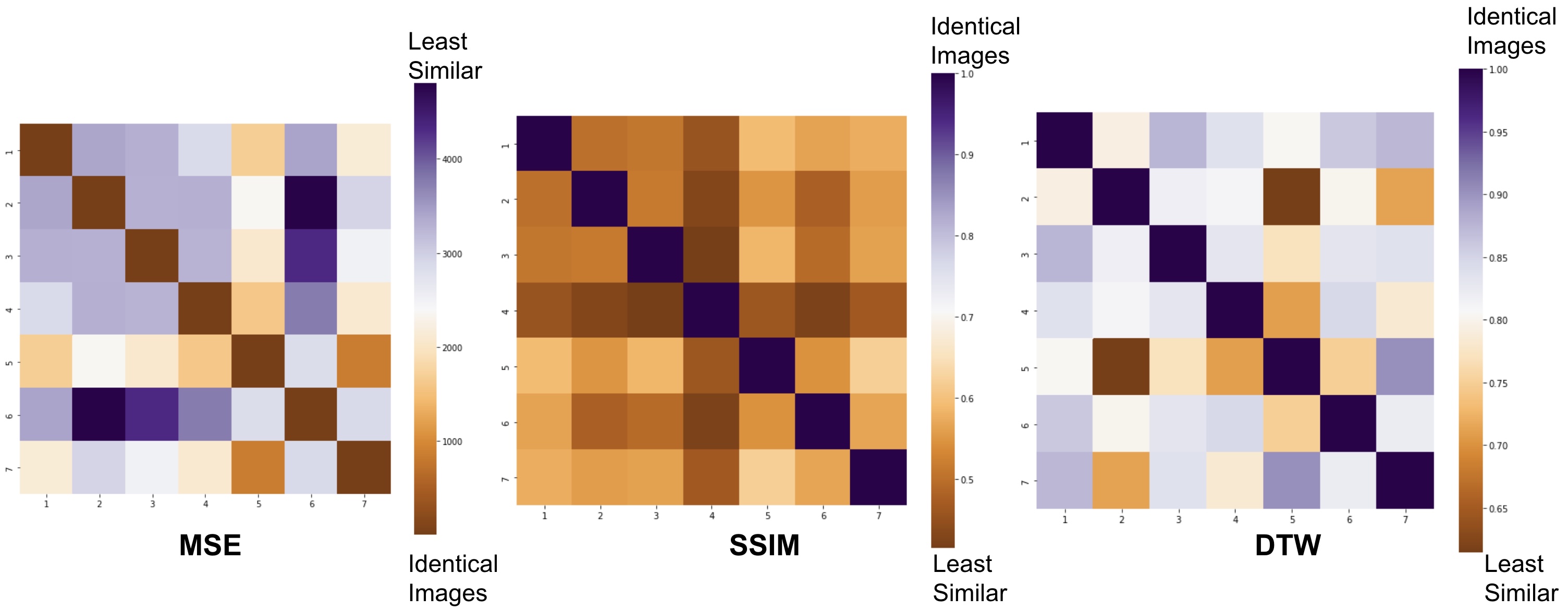}
  \caption{Similarity matrices for the ``of" word snippets from Figure~\ref{of_images} using MSE, SSIM, and DTW respectively. \label{fig:MSE} \label{fig:SSIM} \label{fig:DTW}}
\end{figure}

Figure~\ref{fig:DTW} shows the similarity matrix generated for the 7 ``of" word images from Figure~\ref{of_images} and their respective MSE, SSIM and DTW scores for all  pairs, as an illustration of how we process documents with multiple occurrences of the same word.  For the MSE matrix, the sixth image in the list shows highest  differences when compared against each of the others suggesting it is an outlier. For rest of the images, they are slightly similar to each other given the moderate MSE scores.
For SSIM it is the fourth image that shows least similarity with others while for 
DTW the fifth images is most  dissimilar to the others. 

\begin{table}[!htb]
\centering
\caption{Table of similarity values for the ``of" word snippets.}
\label{tab:sims}
        \begin{tabular}{crrr}
        \toprule 
        ``of'' word & MSE & SSIM & DTW\\
        \midrule 
        1  & 0 & 1.00 & 1.00 \\
        2 & 1807 & 0.39 & 0.79 \\
        3 & 1137 & 0.39 & 0.87 \\
        4 & 1028 & 0.37 & 0.83 \\
        5 & 797 & 0.48 & 0.80 \\
        6 & 903 & 0.49 & 0.86 \\
        7 & 746 & 0.48 & 0.87 \\
        \bottomrule
        \end{tabular}
\end{table}

Table~\ref{tab:sims} shows a comparison of the first  (``of'') image to each of the other ``of'' images and  the scores for MSE, SSIM and DTW.  MSE gives unbounded error scores for every difference calculation, whereas SSIM and DTW gives values  in the range  0 to 1. The scores given by DTW are slightly higher than SSIM.  We applied partition-based clustering to each group of instances of each of the 5 target words (the, of, to, in, a) from each of the diary entries, for each of the weeks in each of the October months in each of the 5 years of the diary entries under study.

\begin{figure}[!t]
  \centering
  \includegraphics[width=\linewidth]{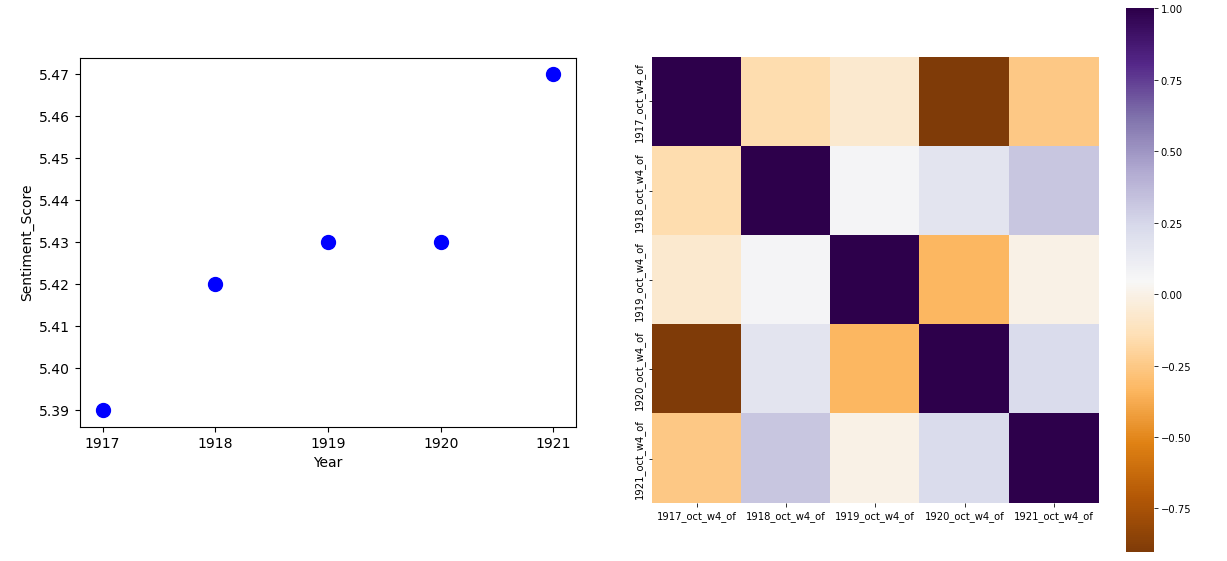}
  \caption{Sample canonical similarity matrix for the word ``of'' for week 4 during October of each year of 1917 to 1921 with graph (left side) illustrating sentiment scores for documents from those weeks.\label{fig:canonical}}
\end{figure}

We then applied spectral clustering to the image data, which considers the task of grouping data as a graph partitioning problem \cite{von2007tutorial}. We specified the number of clusters to be generated as $k=2$ to separate outlier word images from canonical forms, for  each diary entry. For each clustering, we consider the largest cluster of word images as the canonical form for that word in that diary entry. Specifically, this canonical form is the average blended form of all similar-looking word images assigned to one cluster.
We performed similarity comparison among the generated canonical forms for the 5 words in the same  way as was done on the individual word images using the similarity measures. We subsequently created heatmaps for each measure and a comparison table to show how similar or dissimilar the blended versions of the word images are with one another. A sample canonical similarity matrix is shown in Figure~\ref{fig:canonical} relating to different writings of the word ``of'' over  5 years with the sentiment scores for those weeks also shown.

\section{Sentiment Analysis of Diary Entries}
\label{sec:sentiment}

Human language  evolves over time and, as we use words to convey our emotions,  these can display a positive or a negative sentiment. Many open source and commercial tools can now assess the polarity of sentiment in a given text input, although the effectiveness can vary from one tool to another  \cite{10.1145/3532203}.
One popular system is the Hedonometer \cite{10.1073/pnas.1411678112},  a vocabulary-based sentiment analysis tool which computes a ``happiness score'' for text. Here, a list of c.10,000 words have been manually assigned or tagged with a sentiment score which ranges from 1 to 9 (1=sad, 5=neutral, 9=happy) and the sentiment of a document is an aggregation of the scores of its individual words.
The Hedonometer  has previously been applied to a variety of different text types \cite{frank2013happiness}
Since the text in the Jacob diaries is around 100 years old, using a standard commercial sentiment analysis tool could generate unreliable results  given the slightly different vocabularies and language use which might appear in modern vs. older texts. Thus we use Hedonometer scores to calculate an equivalent of sentiment scores for the Jacob diary entries.

\begin{figure*}[!htb]
\includegraphics[width=0.8\linewidth]{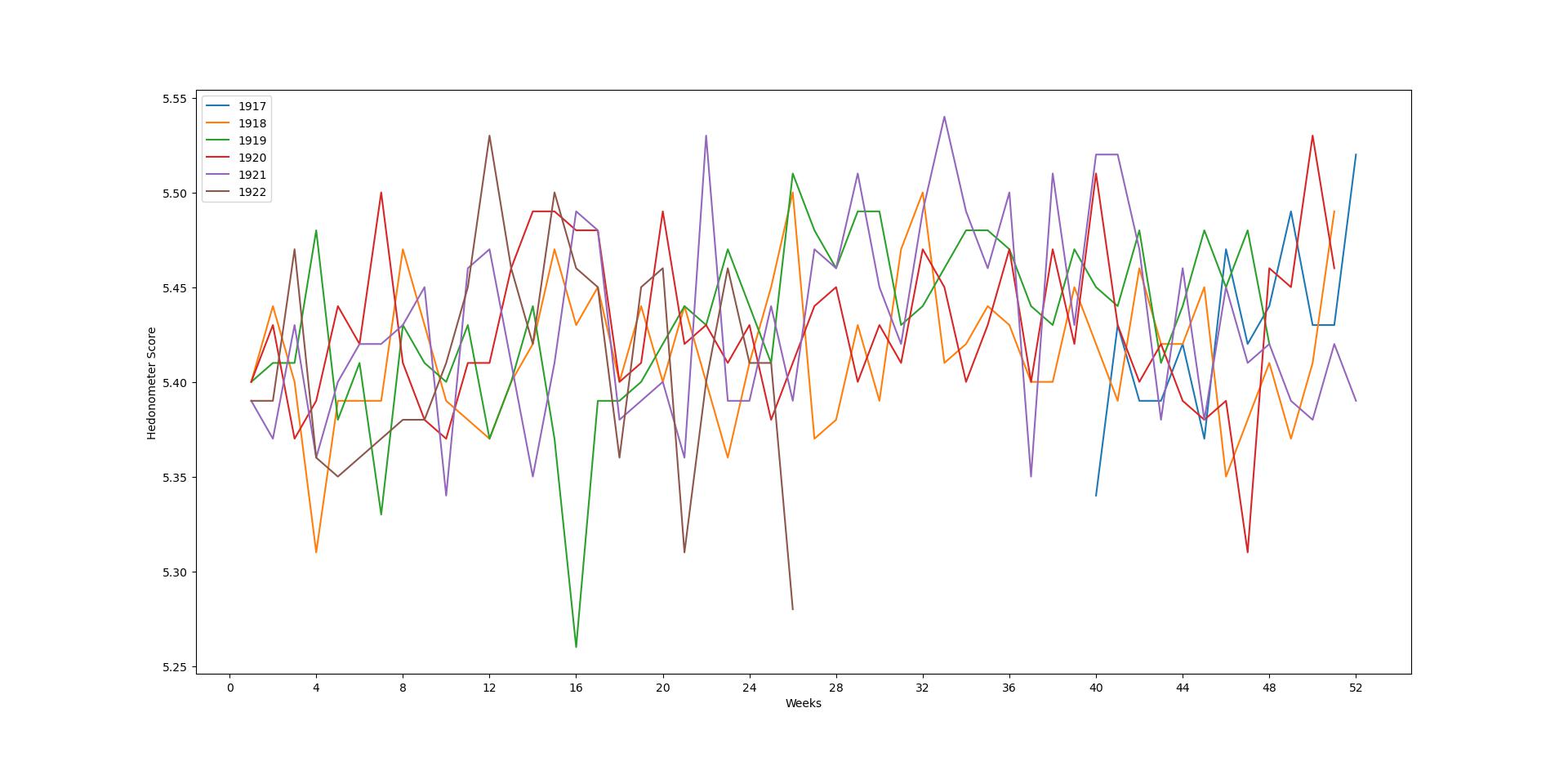}
\vskip -1.3em
\caption{Sentiment/Hedonometer scores for the Jacob diaries computed on a weekly basis over a 5 year period.\label{hedonometer}}
\end{figure*}

Hedonometer sentiment scores  were calculated for each week in each year  in a 5-year period and shown in Figure~\ref{hedonometer}. This  suggests  there is no strong seasonality visible with respect to  sentiment scores across the years. This is interesting given the prominence of walking long distances in Jacob's diaries, both as her primary mode of transport  and as a hobby. The intensity of political events and her involvement in them during this period may explain this lack of seasonal variation.  However, there are some persuasive indications that the Hedonometer scores reflect the impact of external events. For example, one of the lowest points on the Hedonometer scale coincides with a period in late April 1919 where she  heard reports of sexualised violence against young women perceived as collaborators, including an incident in Limerick where `young men there  caught 6 girls that had been walking with soldiers and cut their hair off for a punishment'\footnote{See diary entry MS352/35, p.150}. This is also a period when she listens at length to Hannah Sheehy Skeffington's accounts of the murder of the latter's husband in 1916.  Hedonometer analysis identifies a similar point of deeply negative sentiment immediately preceding the outbreak of the Irish Civil War in June 1922. More significant than these individual points of correlation, however, is the overall volatility of sentiment identified, marked by quite erratic highs and lows. While this may be influenced by the political volatility of the times, it is also indicative of a life lived with great intensity. Jacob was a woman of strong beliefs and passionate personal and political attachments.

With this in mind, we now examine handwriting variations with respect to the sentiment of the topic that was being written as the subject of a diary entry by  comparing the multiple canonical forms of  handwritten images for our five key words with the sentiment scores. The rationale for this is that 
at times during our lives, our experiences and the  emotions we feel will vary depending on the people we spend our time with. When we document these experiences, that documentation and the way we write it may reflect our sentiments about the person or the place or the event we are writing about. This is especially true in any form of journaling and provides an important context for interpreting the results of sentiment analysis.

We illustrate this handwriting analysis  using ``Dorothy"\footnote{Dorothy Macardle was another Irish republican who at one point shared a jail cell with Jacob}  as a  named entity and we selected all  weekly diaries entries in the period 1917--22 in which this entity was mentioned. While we focus on this example the same technique could be applied to occurrences of any named entity. 
To locate  occurrences of ``Dorothy" in the diary images, we used an automatic named entity recognition model. The output from this  includes the matching percentage of the entity and the location where the  entity was mentioned. This work\footnote{We thank Naushad Alam and Jiang Zhang for their help with the entity recognition model and OCR on the transcripts.}
involved resolving multiple surface forms for a  named entity. For example, there are times when Jacob mentioned  ``Tramore"  as  ``Trá Mór" (a location in Ireland), so we were able to find all the occurrences of  ``Tramore" even when there were multiple surface forms of the written word. Using  entity recognition  we used the correct diary images  as  input to  handwriting analysis. 

\begin{figure*}[ht]
\centering
\includegraphics[width=\linewidth]{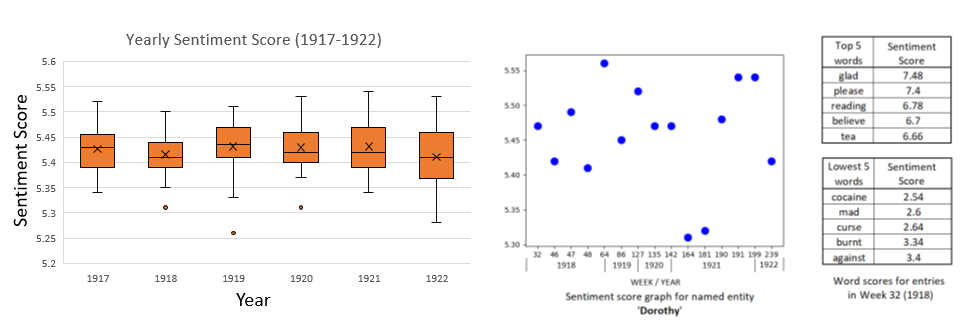}
\vskip -0.5em
\caption{Sentiment scores for diary entries containing the named entity `Dorothy' and  corresponding Hedonometer scores.\label{hedonometer_named_entities}}
\end{figure*}

There were 15 occurrences of the entity  ``Dorothy" in the transcripts for the period 1917-22 and these transcripts were used as an input to  Hedonometer sentiment analysis  to calculate a  score for each of those entries. In calculating the Hedonometer scores, named entities such as ``Dorothy" do not occur in the Hedonometer vocabulary and are regarded as out of vocabulary, and do not contribute to the sscore for a document.  Figure~\ref{hedonometer_named_entities} shows the sentiment scores for those diary entries.  The box-plot summarises the  yearly sentiment scores for all  diary entries while the scatter-plot represents the sentiment scores for diary entries with the named entity ``Dorothy".  To provide  context for the Hedonometer  scores for those diary entries, the tables besides the plot lists the top five  most positive and top five most negative words  present in the diary entry for a specific week -- week 32 of the year 1918. The number of weeks on the x-axis are the relevant week numbers in the transcripts for the period Octocber 1917 to July 1922.

From Figure~\ref{hedonometer_named_entities} we  see  Hedonometer sentiment scores for all diary entries for all years is approximately 5.42. The scores for the 15 entries  making reference to Dorothy are higher than the average, with two standout entries in 1921 having scores between 5.30 and 5.35.  For all other diary entries, their scores are significantly higher than the average scores for their year. 

These inconsistent results are congruent with observations made by Jacob’s biographer, Leeann Lane.  While Jacob was highly suspicious of the role of emotion in politics, she was prone to  strong fluctuations in her friendships and alliances in her private life \cite{lane2011rosamond}.
These findings also point towards another conclusion relevant to the use of sentiment analysis of historical sources more broadly. Jacob’s life writing is influenced by the tradition of reflective and meditative journaling embedded in her Quaker background, but she herself would later become a novelist and historian and she uses language for very deliberate effects. She had a lively, witty  and sometimes satiric writing style in her diaries, characterised by the use of very definitive  modifiers  to describe people and  places. The satirical impulse in her social and personal observations tends to give them a negative bias.  This is undoubtedly a factor in the patterns detected by sentiment analysis. Rosamond Jacob’s  distinctive literary style contributes to the intensity and volatility of positive and negative sentiment in her diaries. 

\section{Conclusions}
\label{sec:conc}
\label{sec:disc}
We  present a methodology  describing how we processed a segment of 5 years of  digitised diary entries of Rosamond Jacob in a systematic manner. 
For handwriting analysis, we found  the technique of Dynamic Time Warping consistently captures more similarity among instances of the same, commonly occurring handwritten words than does MSE or  SSIM. For this reason we used DTW to generate canonical forms of 5 commonly occurring words  for each diary entry.  
For  sentiment analysis of  transcripts of diary entries, we used a vocabulary-based approach  based on a dictionary of c.10,000 annotated words excluding named entities, rather than building or using a contemporary off-the-shelf sentiment analysis tool which might have struggled because of mismatches in vocabulary or word use given the age of the diary. We used Hedonometer scores to build sentiment graphs for weekly diary entries and diary entries around named entities.

In linking handwriting analysis to mood analysis for the October diary entries, Figure~\ref{hedonometer_named_entities} shows a sentiment graph with the range of sentiment scores  for each year. While the range of scores in the years 1919 and  1920 are almost same, their canonical forms do not show a high level of similarity. This might be attributed to the paper quality of the digitised collection, where some diary entries had faint ink and low brightness which affected the quality of the image segmentation. 

The results reported here are applied to a limited segment of 5 years of diary entries, we have shown that there is scope to work with a larger time range of diary entries and this is our future work. We have also shown there is potential to focus on  handwriting changes for diary entries around a given topic and we have illustrated this for diary entries around the named entity ``Dorothy''.   
The work to date has not establish any strong correlation between  handwriting and  sentiment behind the handwritten words. 
Further work  has considerable potential to illuminate the ways in which  handwriting and sentiment analysis of life writing can enable new forms of historical and cultural research. From an analysis perspective we will also work to determine the applicability of the hedonometer vocabulary to texts which are more than a century old.

\subsubsection*{Acknowledgements}
This research was conducted with  financial support of Science Foundation Ireland [12/RC/2289\_P2] at Insight the SFI Research Centre for Data Analytics at Dublin City University.

\bibliographystyle{plain} 
\bibliography{bibliography.bib} 

\end{document}